\documentclass[a4paper,11pt]{article} 

\usepackage{amsmath,amssymb}
\usepackage{tabularx,epsfig,color,tabularx,xcolor,comment}
\usepackage{algorithm}
\usepackage{algpseudocode}
\usepackage{fullpage}
\usepackage{bm}
\usepackage{dcolumn}
\usepackage{authblk}
\usepackage[colorlinks,urlcolor=red,citecolor=red,linkcolor=blue,pdftex, pdfauthor = Lukas Exl]{hyperref}

\usepackage{lineno,hyperref}
\modulolinenumbers[5]

\title{\Large{\textbf{Preconditioned nonlinear conjugate gradient method for micromagnetic energy minimization}}}
\author[1,2]{Lukas Exl\thanks{\texttt{lukas.exl@univie.ac.at}}} 
\affil[1]{Faculty of Mathematics, University of Vienna, Oskar-Morgenstern-Platz 1, 1090 Wien, Austria.} 
\affil[2]{Institute for Analysis and Scientific Computing, Vienna University of Technology, Wiedner Hauptstra{\ss}e 8-10, 1040, Vienna.} 
\author[3]{Johann Fischbacher\thanks{\texttt{johann.fischbacher@donau-uni.ac.at}}}
\author[3]{Alexander Kovacs\thanks{\texttt{alexander.kovacs@donau-uni.ac.at}}}
\author[3]{Harald Oezelt\thanks{\texttt{harald.oezelt@donau-uni.ac.at}}}
\author[3]{Markus Gusenbauer\thanks{\texttt{markus.gusenbauer@donau-uni.ac.at}}}
\author[3]{Thomas Schrefl\thanks{\texttt{tschrefl@gmail.com}}}
\affil[3]{Center for Integrated Sensor Systems, Danube University Krems, Viktor Kaplan Str. 2/E, 2700 Wiener Neustadt, Austria}

\begin{document}
%
\maketitle
\date

\noindent\textbf{Abstract.} Fast computation of demagnetization curves is essential for the computational design of soft magnetic sensors or permanent magnet materials. 
We show that a sparse preconditioner for a nonlinear conjugate gradient energy minimizer can lead to a speed up by a factor of 3 and 7 for computing  hysteresis in soft magnetic and hard magnetic materials, respectively. 
As a preconditioner an approximation of the Hessian of the Lagrangian is used, which only takes local field terms into account.  
Preconditioning requires a few additional sparse matrix vector multiplications per iteration of the nonlinear conjugate gradient method, which is used for minimizing the energy for a given external field. 
The time to solution for computing the demagnetization curve scales almost linearly with problem size. \\


\section{Introduction}

Micromagnetic simulations have become an essential tool for the design of soft magnetic sensors \cite{brueckl2017vortex} and permanent magnets \cite{bance2014high, sodervznik2017magnetization}. 
Fast computation of the demagnetization curve is a prerequisite for parameter studies which are applied to find optimal designs \cite{kovacs2017micromagnetic}. 
Micromagnetic models which are used for the analysis and design of permanent magnets are huge. Typically the number of finite elements exceeds 20 million elements \cite{sepehri2016micromagnetic, oikawa2016large}. As a consequence fast solvers for computing the demagnetization curve are required. The computation of the demagnetization curve by minimization of the total energy  for subsequent external fields \cite{exl2014labonte, acevedo2016efficient, tanaka2017speeding} has proven to be faster than the solution of the dynamical system according to the Landau-Lifshitz  Gilbert equation. 
Micromagnetic solvers based on energy minimization are also the key building blocks for computing the minimum energy path with the string method \cite{bance2014high, acevedo2016efficient, fischbacher2017limits}. This method gives the barrier height for thermally activated switching and thus is used to compute the reduction of the coercive field by thermal fluctuations.

The derivative and function evaluations in dynamical and static micromagnetic computations are very expensive, mostly due to the nonlocal stray field component. 
However, the use of a stray field algorithm that scales optimally or quasi-optimially with problem size such as the algebraic multigrid method \cite{sun2006adaptive,shepherd2014discretization,exl2014labonte}, the fast multipole method \cite{yuan1992fast,apalkov2003fast,palmesi2017high} and hierarchical matrices \cite{popovic2005applications}, fast Fourier transform based methods (FFT) \cite{yuan1992fast,fabian1996three,tsukahara2016implementation} or non-uniform FFT algorithms \cite{kritsikis2014beyond,exl2014non,fu2015micromagnetics} is not sufficient to obtain a micromagnetic solver that scales linearly with problem size.  Owing to the exchange interactions the micromagnetic equations can be considered as stiff \cite{della1993stiff} and the number of time steps in a dynamical solver or the number of iterations in a static solver increase with increasing problem size. 
Stiffness typically leads to CFL-type step size restrictions of the form $\Delta t \leq C \,\Delta x^2$ where $\Delta t$ is the time step size and $\Delta x$ is the mesh size. When the number of steps increases with a finer mesh (or larger problem size) the complexity of the overall algorithm is above linear. 

This problem has been treated successfully in dynamic micromagnetics. Higher order explicit integrators are an alternative if extensive evaluation of the nonlocal field component is avoided \cite{exl2017extrapolated}. On the other hand, (semi-) implicit schemes try to tackle the stiffness directly at the expense of additionally having to solve (linear) systems. Preconditioning these systems is mandatory, where again stray field computations have to be avoided. We stress that potential preconditioners are often built up by local field contributions, which might lead to linear systems for evaluation of the  preconditioner with high condition numbers due to similar reasons as those leading to the above step size restriction. However, practical tests show that rough approximations of the solution of such preconditioning systems suffice. Suess et al. \cite{suess2002time} and Sheperd et al. \cite{shepherd2014discretization}  investigated the  correlation between the microstructure of the magnet  and the discretization on the stiffness of the Landau-Lifshitz Gilbert equation. They show, that implicit or semi-implicit integrators of the Landau-Lifshitz equation with a preconditioned linear system for the Newton step, successfully tackles stiffness in micromagnetic systems. As preconditioner the sparse part of the Jacobian matrix, which is related to the exchange and anisotropy energy, was used. We will orientate our approach for preconditioning the nonlinear conjugate gradient method for energy minimization to these findings for the dynamical system case and will show that a static micromagnetic solver with optimal scaling can be achieved by proper preconditioning.

In micromagnetics several energy minimization methods have been applied. Accelerated steepest descent methods, which were originally developed for color image denoising \cite{goldfarb2009curvilinear}, 
were used for the simulation of permanent magnets with a finite element scheme \cite{exl2014labonte} and for the simulation of sensor elements with finite difference \cite{abert2014efficient}.
The conjugate gradient method was used for the simulation of soft magnetic elements \cite{koehler1992finite} and permanent magnets \cite{fischbacher2017nonlinear, tanaka2017speeding}. 
Garcia-Cervera \cite{garcia2008numerical} and Escobar \cite{acevedo2016efficient} discussed a truncated Newton method for minimizing the micromagnetic energy. 
They also propose to use a sparse part of the Hessian matrix for building a preconditioner for solving the linear system at each Newton step. 
Escobar  \cite{acevedo2016efficient} reported a reduction of the CPU time by a factor 1/3 when using a preconditioner that takes into account the local interactions only. 
On the other hand no preconditioners have been applied for the nonlinear conjugate gradient method in micromagnetics yet. 
The success of the preconditioned Newton method by Escobar  \cite{acevedo2016efficient} suggests that a similar preconditioner might be successful for nonlinear conjugate gradient methods too.  

Hysteresis in a non-linear system, like a permanent magnet, results from the path formed by subsequently following local minima in an energy landscape progressively changed by a 
varying external field \cite{kinderlehrer1994simulation}. The numerical experiments by Fischbacher et al. \cite{fischbacher2017nonlinear} show that 
particular care has to be taken when energy minimization is used to compute hysteresis. 
The minimizer may fail to reach the closest local minimum. If local minima are missed by the algorithm the demagnetization curve shows only one or few large steps and the distinct switching fields of different grains will not be resolved.
This failure can be avoided by carefully choosing the search direction and the step length for the numerical energy minimization method. 
This can be achieved using a modified Hestensen-Stiefel method \cite{hestenes1952methods}, whereas Quasi-Newton methods with projected updates failed to produce the correct demagnetization curve \cite{fischbacher2017nonlinear}.
In this work we present a preconditioned nonlinear conjugate gradient solver for micromagnetics and show how the method specific parameters can be tuned for computing hysteresis in soft magnetic elements or hard magnetic materials.  

\section{Method}

\subsection{Conjugate gradient methods}

We first briefly review the preconditioned (nonlinear) conjugate gradient method. Let us start with the  preconditioned linear conjugate gradient algorithm.
It was originally introduced by Hestenes and Stiefel \cite{hestenes1952methods} for the solution of a
linear system of equation $\mathbf{Cx} = -\mathbf{b}$ where $\mathbf{C}$ is symmetric and positive definite (SPD), which is equivalent to
minimizing the quadratic function 
\begin{equation}
\label{eq:quad}
Q(\mathbf{x}) = \frac{1}{2}\mathbf{x}^{\mathrm T}\mathbf{Cx} + \mathbf{b}^{\mathrm T}\mathbf{x} + \mathbf{c},
\end{equation}
with respect to the vector $\mathbf{x}$. The preconditioned linear conjugate gradient method is given by Algorithm \ref{alg1} \cite{nocedal2006numerical}.
\begin{algorithm}\caption{Preconditioned linear conjugate gradient method}
\label{alg1}
Task: Solve $\mathbf{Cx} = -\mathbf{b}$ or minimize $\tfrac{1}{2}\mathbf{x}^{\mathrm T}\mathbf{Cx} + \mathbf{b}^{\mathrm T}\mathbf{x} + \mathbf{c}$.\\
Initialize: 
\begin{algorithmic} 
 \State Start with an initial guess $\mathbf{x}_{0}$ and compute the residual $\mathbf{g}_{0} = \nabla Q(\mathbf{x}_{0}) = \mathbf{Cx} + \mathbf{b}$.
 \State Solve ${\mathbf{P}} \mathbf{y}_{0} = \mathbf{g}_{0}$ ($\mathbf{P}$ is a SPD preconditioner)
 \State Set the initial search direction $\mathbf{d}_{0} = -\mathbf{y}_{0} $
\end{algorithmic}
Minimize:
\begin{algorithmic}
 \For{$j= 0,1,2,\hdots$}
 \State Compute the step length   $\alpha_j = \frac{\mathbf{g}_{j}^{\mathrm T}\mathbf{y}_{j}}{\mathbf{d}_{j}^{\mathrm T} \mathbf{C}\mathbf{d}_{j}}$ 
 \State Set new solution vector   $\mathbf{x}_{j+1} = \mathbf{x}_{j} + \alpha_{j} \mathbf{d}_{j}$ 
 \State Update the residual       $\mathbf{g}_{j+1} = \mathbf{g}_{j} + \alpha_{j} \mathbf{C} \mathbf{d}_{j}$ 
 \State Exit if                   $|\mathbf{g}_{j+1}| < \epsilon$ (for some $\epsilon >0$)
 \State Solve                     $\mathbf{P} \mathbf{y}_{j+1} = \mathbf{g}_{j+1}$ 
 \State Compute                   $\beta = \frac{\mathbf{g}_{j+1}^{\mathrm T} \mathbf{y}_{j+1}}{\mathbf{g}_{j}^{\mathrm T} \mathbf{y}_{j}}$ 
 \State Compute search direction  $\mathbf{d}_{j+1} = - \mathbf{y}_{j+1} + \beta  \mathbf{d}_{j}$ 
 \EndFor
\end{algorithmic}
\end{algorithm}

The preconditioner $\mathbf{P}$ is a symmetric and positive definite approximation of $\mathbf{C}$ so that $\mathbf{P} \mathbf{y} = \mathbf{g}$ is more easily solved than $\mathbf{C} \mathbf{y} = \mathbf{g}$. 
The aim of preconditioning is to reduce the number of iterations until convergence. This can be achieved if  
the spectrum of the system matrix of the preconditioned system, $\mathbf{P}^{-1}\mathbf{C}$, is changed such that eigenvalues are clustered or the condition number is improved.

If the function to be minimized, $F(\mathbf{x})$, is not quadratic we can apply the nonlinear conjugate gradient method.  It is given in Algorithm \ref{alg2} \cite{pytlak2008conjugate}. 
The solution vector $\mathbf{x}$ is updated iteratively, until convergence is reached. 

\begin{algorithm}\caption{Nonlinear conjugate gradient method}
\label{alg2}
Task: Minimize $F(\mathbf{x})$.\\
Initialize: 
\begin{algorithmic} 
 \State Start with an initial guess $\mathbf{x}_{0}$ and compute the energy gradient $\mathbf{g}_{0} = \nabla F(\mathbf{x}_{0})$. 
 \State Set the initial search direction $\mathbf{d}_{0} = -\mathbf{g}_{0}$
\end{algorithmic}
Minimize:
\begin{algorithmic}
 \For{$j= 0,1,2,\hdots$}
 \State Compute the step length  $\alpha_j$ by minimizing  $F(\mathbf{x}_{j}+\alpha_{j}\mathbf{d}_{j})$ (line search)
 \State Set new solution vector  $\mathbf{x}_{j+1} = \mathbf{x}_{j} + \alpha_{j} \mathbf{d}_{j}$ 
 \State Compute energy gradient  $\mathbf{g}_{j+1}  = \nabla F(\mathbf{x}_{j+1})$ 
 \State Compute search direction $\mathbf{d}_{j+1}  = - \mathbf{g}_{j+1} + \beta  \mathbf{d}_{j}$ 
 \EndFor
\end{algorithmic}
\end{algorithm}

Different ways to compute the value of $\beta$ for the update of the search direction in the non-linear conjugate gradient method were proposed and are given in equations 
(\ref{eq:FR}) to (\ref{eq:HS}). They are equivalent if $F(\mathbf{x})$ is quadratic.
Originally, Fletcher and Reeves \cite{fletcher1964function} proposed
\begin{equation}
\label{eq:FR}
  \beta^{\mathrm{FR}} = \frac{\mathbf{g}_{j+1}^{\mathrm T}\mathbf{g}_{j+1}}{\mathbf{g}_{j}^{\mathrm T}\mathbf{g}_{j}}.
\end{equation}
The Polak-Ribiere-Polyak conjugate gradient \cite{polak1969note,polyak1969conjugate} method uses
\begin{equation}
\label{eq:PRP}
  \beta^{\mathrm{PRP}} = \frac{(\mathbf{g}_{j+1}-\mathbf{g}_{j})^{\mathrm T}\mathbf{g}_{j+1}}{\mathbf{g}_{j}^{\mathrm T}\mathbf{g}_{j}}.
\end{equation}
Nonlinear conjugate gradient methods that use (\ref{eq:PRP}) instead of  (\ref{eq:FR}) are believed to be more efficient than nonlinear conjugate gradient algorithms that
use (\ref{eq:FR}), because of the self-correcting behavior. When 
$|\mathbf{g}_{j+1}-\mathbf{g}_{j}|$ is small, $\beta$ is close to zero, and the search direction 
is effectively $\mathbf{d}_{i+1} = - \mathbf{g}_{i+1}$ which results in a restart of the iterations. 
The choice
\begin{equation}
\label{eq:HS}
  \beta^{\mathrm{HS}}  = \frac{(\mathbf{g}_{j+1}-\mathbf{g}_{j})^{\mathrm T}\mathbf{g}_{j+1}}
                                 {(\mathbf{g}_{j+1}-\mathbf{g}_{j})^{\mathrm T}\mathbf{d}_{j}}
\end{equation}
goes back to Hestensen and Stiefel \cite{hestenes1952methods}. It is equivalent to (\ref{eq:PRP}) if the line search is exact. The search directions of a nonlinear conjugate gradient
method that use (\ref{eq:HS}) fulfill the conjugacy condition
\begin{equation}
  (\mathbf{g}_{j+1}-\mathbf{g}_{j})^{\mathrm T}\mathbf{d}_{j+1} = 0
\end{equation}
independent of the accuracy of the line search. This property turned out to be helpful in micromagnetics. For computing demagnetization curves only an approximate line search \cite{fischbacher2017nonlinear}, which avoids too long steps, can be used in order to obtain correct results.

We now can assume that $F(\mathbf{x})$ is quadratic in the vicinity of $\mathbf{x}_{j}$ (the current iterate) and hope that a transformation of the problem 
near $\mathbf{x}_{j}$ speeds up convergence similar to preconditioning of the linear conjugate gradient method. The preconditioned nonlinear conjugate gradient method \cite{pytlak2008conjugate} is given in Algorithm \ref{alg3}.

\begin{algorithm}\caption{Preconditioned nonlinear conjugate gradient method}
\label{alg3}
Task: Minimize $F(\mathbf{x})$.\\
Initialize: 
\begin{algorithmic} 
 \State Start with an initial guess $\mathbf{x}_{0}$ and compute the energy gradient $\mathbf{g}_{0} = \nabla F(\mathbf{x}_{0})$. 
 \State Solve ${\mathbf{P}}\mathbf{y}_{0} = \mathbf{g}_{0}$ 
 \State Set the initial search direction $\mathbf{d}_{0} = -\mathbf{y}_{0}$
\end{algorithmic}
Minimize:
\begin{algorithmic}
 \For{$j= 0,1,2,\hdots$}
 \State Compute the step length $\alpha_j$ by minimizing  $F(\mathbf{x}_{j}+\alpha_{j}\mathbf{d}_{j})$ (line search)
 \State Set new solution vector  $\mathbf{x}_{j+1} = \mathbf{x}_{j} + \alpha_{j} \mathbf{d}_{j}$ 
 \State Compute energy gradient  $\mathbf{g}_{j+1}  = \nabla F(\mathbf{x}_{j+1})$ 
 \State Solve                    $\mathbf{P} \mathbf{y}_{j+1} = \mathbf{g}_{j+1}$ 
 \State Compute search direction $\mathbf{d}_{j+1}  = - \mathbf{y}_{j+1} + \beta^{*}  \mathbf{d}_{j}$  
 \EndFor
\end{algorithmic}
\end{algorithm}

The matrix $\mathbf{P}$ is an approximation of the Hessian of $F$. Again we assume that $\mathbf{P} \mathbf{y} = \mathbf{g}$ can be easily solved.
The equation $\eqref{eq:FR}$ to $\eqref{eq:HS}$ have to be modified as follows \cite{andrea2017novel}:

\begin{eqnarray}
\label{eq:FRstar}
  \beta^{\mathrm{*FR}}& = &\frac{\mathbf{g}_{j+1}^{\mathrm T}\mathbf{y}_{j+1}}{\mathbf{g}_{j}^{\mathrm T}\mathbf{y}_{j}}, \\
\label{eq:PRPstar}
  \beta^{\mathrm{*PRP}}& = &\frac{(\mathbf{g}_{j+1}-\mathbf{g}_{j})^{\mathrm T}\mathbf{y}_{j+1}}{\mathbf{g}_{j}^{\mathrm T}\mathbf{y}_{j}}, \\
\label{eq:HSstar}
  \beta^{\mathrm{*HS}} & = &\frac{(\mathbf{g}_{j+1}-\mathbf{g}_{j})^{\mathrm T}\mathbf{y}_{j+1}}
                                 {(\mathbf{g}_{j+1}-\mathbf{g}_{j})^{\mathrm T}\mathbf{d}_{j}}.
\end{eqnarray}
We want to stress that preconditioning a nonlinear conjugate gradient method by a sequence of symmetric and positive matrices that are approximations to the Hessian of the objective does not have to lead to 
improvements in convergence at all. For instance, the preconditioned nonlinear conjugate gradient method with exact Hessian at the current iterate as preconditioners can be seen as a combination of the Newton 
and conjugate gradient method, where the analysis in Ref.\cite{pytlak2008conjugate} shows no improvement over the Newton method. In fact, convergence with $Q$-order greater than one is not guaranteed. However, preconditioned 
nonlinear conjugate gradient algorithms mimic a linear conjugate gradient method applied to a strongly convex quadratic approximation of the objective (in a neighborhood of a solution). 
Instead of fixing a preconditioner matrix beforehand, we change it each or every fixed number of iteration to account for the fact that the quadratic model changes as well.

\subsection{Micromagnetics energy minimization}

After discretization by the finite element method or the finite difference method the micromagnetic energy can be written as a quadratic form \cite{fischbacher2017nonlinear}:
\begin{equation}
\label{eq:energy}
F(\mathbf m) = \frac{1}{2} \mathbf{m}^{\mathrm T} \mathbf{C} \mathbf{m} - \frac{1}{2}\mathbf{h}_{\mathrm d}^{\mathrm T} \mathbf{M} \mathbf{m} - \mathbf{h}_{\mathrm{ext}}^{\mathrm T} \mathbf{M} \mathbf{m}. 
\end{equation}
Here, we assume the mesh-vector $\mathbf{m} \in \mathbb{R}^{3N}$ and will denote associated 3-vectors at node $i$ with $\mathbf{m}^{(i)},\,i=1,\hdots,N$. 
The first, second, and last term of the right hand side of (\ref{eq:energy}) are the sum of the exchange and anisotropy
energy, the magnetostatic self energy, and the Zeeman energy, respectively. The sparse matrix $\mathbf C$ contains
grid information associated with the exchange and anisotropy energy. The matrix $\mathbf M$ accounts for the local variation of the saturation magnetization $M_\mathrm{s}$ within the magnet \cite{schrefl_numerical_nodate}.
The vectors $\mathbf m$, $\mathbf{h}_{\mathrm d}(\mathbf{m})$, and $\mathbf{h}_{\mathrm ext}$ collect the unit vector of the magnetization, the demagnetizing field (which is linear in $\mathbf{m}$), and the external
field at the nodes of the finite element mesh or the cells of a finite difference grid, respectively. We denote the subvectors associated with a grid point with the superscript $(i)$. 
For example $\mathbf{m}^{(i)}$ is the unit vector parallel to the magnetization at point $i$. We can split the matrix $\mathbf{C} = \mathbf{C}_{A} + \mathbf{C}_{K}$ into the
contribution from the ferromagnetic exchange energy, $\tfrac{1}{2} \mathbf{m}^{\mathrm T} \mathbf{C}_{A} \mathbf{m}$, and the contribution from the magnetocrystalline anisotropy energy, 
$\tfrac{1}{2}  \mathbf{m}^{\mathrm T} \mathbf{C}_{K} \mathbf{m}$.
 
In micromagnetics we have to treat the non-convex constraints $|\mathbf{m}^{(i)}|=1$ at the grid points $(i)$, which is also called the \textit{unit norm constraint}. 
We therefore get the non-convex and non-linear optimization problem 
\begin{align}\label{eq:opt}
 \min_{\mathbf{m} \in \mathbb{R}^{3N}} F(\mathbf m) \quad \mathrm{subject\, to} \quad |\mathbf{m}^{(i)}|=1,\, i=1,\dots,N.
\end{align}
Let us follow Ref.\cite{exl2014thesis} and reformulate \eqref{eq:opt} by looking closer at the Lagrangian associated with \eqref{eq:opt}, its gradient and the action of the Hessian. The Lagrangian is
\begin{align}
 \mathcal{L}_F(\mathbf{m}) = F(\mathbf{m}) - \bm{\lambda}^{\mathrm{T}} \mathbf{c}(\mathbf{m}),
\end{align}
where $\bm{\lambda} \in \mathbb{R}^N$ collects the Lagrange multipliers and $\mathbf{c}(\mathbf{m}) \in \mathbb{R}^N$, a constraint violation function with components $c(\mathbf{m})^{(i)} = \tfrac{1}{2}(|\mathbf{m}^{(i)}|^2 - 1)$. There holds 
\begin{align}\label{eq:nablaLF1}
 \nabla_\mathbf{m}  \mathcal{L}_F(\mathbf{m})^{(i)} =  \nabla F(\mathbf{m})^{(i)} - \lambda^{(i)}\mathbf{m}^{(i)}
\end{align}
and the Karush-Kuhn-Tucker (KKT) conditions \cite{nocedal2006numerical} yield
\begin{align}
 \lambda^{(i)} = {\mathbf{m}^{(i)}}^{\mathrm{T}} \nabla F(\mathbf{m})^{(i)}. 
\end{align}
Hence, by substituting this into \eqref{eq:nablaLF1} we get 
\begin{align}\label{eq:nablaLF}
 \nabla_\mathbf{m} \mathcal{L}_F(\mathbf{m}) = \Pi_{\mathbf{m}^{\bot}}( \nabla F(\mathbf{m})),  
\end{align}
with the orthogonal projection $\Pi_{\mathbf{m}^\bot}$ defined as 
\begin{align}\label{eq:proj}
 (\Pi_{\mathbf{m}^\bot} \mathbf{v})^{(i)} = \mathbf{v}^{(i)} - ({\mathbf{v}^{(i)}}^{\mathrm{T}} \mathbf{m}^{(i)}) \mathbf{m}^{(i)}.  
\end{align}
The necessary optimality condition for micromagnetic constrained equilibrium is therefore 
\begin{align}\label{eq:optcond}
 \Pi_{\mathbf{m}^\bot}( \nabla F(\mathbf{m})) = 0.
\end{align}
Note that the objective of the following problem 
\begin{align}\label{eq:opt2}
  \min_{\mathbf{m}} \widetilde{F}(\mathbf m) := F(\mathcal{N}\mathbf{m}) \quad \mathrm{with} \quad   (\mathcal{N}\mathbf{m})^{(i)} = \frac{\mathbf{m}^{(i)}}{|\mathbf{m}^{(i)}|}
\end{align}
has the gradient 
\begin{align}
 \nabla \widetilde{F}(\mathbf m)^{(i)} =  \frac{1}{|\mathbf{m}^{(i)}|} \big( \nabla F(\mathcal{N}\mathbf m)^{(i)} - \frac{{\mathbf{m}^{(i)}}^{\mathrm{T}} \nabla F(\mathcal{N}\mathbf{m})^{(i)}}{|\mathbf{m}^{(i)}|^2} \mathbf{m}^{(i)} \big). 
\end{align}
Minimizing the objective in \eqref{eq:opt2} with a numerical solver that ensures the unit norm of the iterates will therefore approximate the condition \eqref{eq:optcond} and thus a constrained solution to the original micromagnetic problem \eqref{eq:opt} 
\cite{koehler1992finite}. \\  
In the following we will need second derivative information for the purpose of preconditioning. Calculating the derivative of \eqref{eq:nablaLF} gives the following expression for the action of the Hessian on a mesh vector $\mathbf{v}$
\begin{eqnarray}\label{eq:hessact}
\begin{aligned}
 (\nabla^2_\mathbf{m} \mathcal{L}_F(\mathbf{m})\mathbf{v})^{(i)} = & \, \big(\nabla^2_\mathbf{m} F(\mathbf{m})\mathbf{v}\big)^{(i)} - 
 \big( ({\mathbf{v}^{(i)}}^{\mathrm{T}} \nabla F(\mathbf{m})^{(i)})\mathbf{m}^{(i)} + \\ &\,({\mathbf{m}^{(i)}}^{\mathrm{T}} \nabla F(\mathbf{m})^{(i)})\mathbf{v}^{(i)} + 
 ({\mathbf{m}^{(i)}}^{\mathrm{T}} \nabla F(\mathbf{v})^{(i)})\mathbf{m}^{(i)}\big),
\end{aligned}
\end{eqnarray}
derived from extracting the linear terms in $\mathbf{v}^{(i)}$ of the expression $\nabla_\mathbf{m} \mathcal{L}_F(\mathbf{m}+\mathbf{v})^{(i)} - \nabla_\mathbf{m} \mathcal{L}_F(\mathbf{m})^{(i)}$.

For calculation of the molecular orientation of liquid crystals Cohen et al. \cite{cohen1989relaxation} proposed a ''projected'' nonlinear conjugate gradient method with the following modifications: 
Replace the energy gradient with its component perpendicular to the current orientation (which corresponds to the projected gradient \eqref{eq:nablaLF}) and normalize the orientation vector after each update, 
for which we use the shortcut $\mathcal{N}\mathbf{v}$ with $(\mathcal{N}\mathbf{v})^{(i)} = \mathbf{v}^{(i)} / |\mathbf{v}^{(i)}|$.\\
The preconditioned nonlinear conjugate gradient method for micromagnetic energy minimization is summarized in Algorithm \ref{alg4}.

\begin{algorithm}\caption{Preconditioned nonlinear conjugate gradient method for micromagnetics}
\label{alg4}
Task: Minimize $F(\mathbf{x})$ subject to $|\mathbf{m}^{(i)}|=1$.\\
Initialize: 
\begin{algorithmic} 
 \State Start with an initial guess $\mathbf{m}_{0}$ with $|\mathbf{m}_0^{(i)}|=1$ and compute $\mathbf{g}_{0} = {\Pi_{\mathbf{m}^\bot_0}}\left(\nabla F(\mathbf{m}_{0})\right)$.
 \State Solve ${\mathbf{P}_0} \mathbf{y}_{0} = \mathbf{g}_{0}$ where $\mathbf{P}_0$ is the current preconditioner.
 \State Set the initial search direction $\mathbf{d}_{0} = -\mathbf{y}_{0}$ 
\end{algorithmic}
Minimize:
\begin{algorithmic}
 \For{$j= 0,1,2,\hdots$}
 \State Compute the step length $\alpha_j$ by minimizing  $F\left({\mathcal{N}}\left(\mathbf{x}_{j}+\alpha_{j}\mathbf{d}_{j}\right)\right)$ (line search)
 \State Set new solution vector  $\mathbf{x}_{j+1} = {\mathcal{N}}(\mathbf{x}_{j} + \alpha_{j} \mathbf{d}_{j})$ 
 \State Compute energy gradient  $\mathbf{g}_{j+1}  = {\Pi_{\mathbf{x}^\bot_{j+1}}}(\nabla F(\mathbf{x}_{j+1}))$
 \State Solve                    $\mathbf{P}_{j+1} \mathbf{y}_{j+1} = \mathbf{g}_{j+1}$ 
 \State Compute search direction $\mathbf{d}_{j+1}  = - \mathbf{y}_{j+1} + \beta^{*\mathrm{ZA}}  \mathbf{d}_{j}$  
 \EndFor
\end{algorithmic}
\end{algorithm}

Algorithm \ref{alg4} is the Hestenes-Stiefel conjugate gradient method with restarts \cite{salleh2016efficient} according to 
\begin{eqnarray}
\beta^{\mathrm{*ZA}} & = &\frac{(\mathbf{g}_{j+1}-\mathbf{g}_{j})^{\mathrm T}\mathbf{y}_{j+1}}
                                 {(\mathbf{g}_{j+1}-\mathbf{g}_{j})^{\mathrm T}\mathbf{d}_{j}}  \;\;\;\mathrm{if}\;
                                 \mathbf{y}_{j+1}^{\mathrm T}\mathbf{g}_{j+1} > \mathbf{y}_{j+1}^{\mathrm T}\mathbf{g}_{j} \\
\\
\beta^{\mathrm{*ZA}} & = & 0 \;\;\;\mathrm{otherwise}
\end{eqnarray}

In Algorithm \ref{alg4} we need to solve the system  $\mathbf{P} \mathbf{y} = \mathbf{g}$. If we solve this linear system with the preconditioned conjugate gradient method (Algorithm \ref{alg1}), 
there is no need to build the preconditioner $\mathbf{P}$ explicitly. All we need is the vector product $\mathbf{Pv}$. 

Owing to the constraints $|\mathbf{m}^{(i)}|=1$ the Hessian, $ \nabla^2_\mathbf{m} \mathcal{L}_F(\mathbf{m})$ of \eqref{eq:nablaLF}, depends on the vector field $\mathbf{m}$. 
A preconditioner should be an approximate Hessian whose re-computation at each iteration is relatively cheap.
For the approximate Hessian vector product 
only the short range interactions in \eqref{eq:hessact} are taken into account \cite{tsiantos2006fast,acevedo2016efficient} (if not previously computed and thus available), that is 
\begin{eqnarray}\label{eq:precond}
  \label{eq:Hv}
  \left( \mathbf{H}(\mathbf{m})\mathbf{v} \right)^{(i)}  & \approx (\mathbf{Cv})^{(i)} & - ({\mathbf{m}^{(i)}}^\mathrm{T}(\mathbf{Cv})^{(i)} )\mathbf{m}^{(i)} \\ \nonumber
                                                                                    && - ({\mathbf{m}^{(i)}}^\mathrm{T}\nabla F(\mathbf{m})^{(i)} )\mathbf{v}^{(i)} \\ \nonumber
                                                                                    && - ({\mathbf{v}^{(i)}}^\mathrm{T}\nabla F(\mathbf{m})^{(i)})\mathbf{m}^{(i)},
\end{eqnarray} 
which yields possible candidates for preconditioners. We found that \eqref{eq:precond} without the last term on the right hand side gave the best performance in our numerical tests.  
Note that a preconditioner should be symmetric and positive definite to ensure descent. Since, the preconditioner system is only solved approximately we need to  
replace $\mathbf{y}_{k}$ by $\mathbf{g}_k$ if $\mathbf{g}_j^\mathrm{T} \mathbf{d}_j >0$.

Note that only $\mathbf{Pv}$ 
is evaluated during the iterative solution of $\mathbf{P} \mathbf{y} = 
\mathbf{g}$ in Algorithm \ref{alg4} by using Algorithm \ref{alg1}.  As preconditioner in 
Algorithm \ref{alg1} we use the diagonal matrix which is built from the exchange interactions, $P_{ij} = \delta_{ij}\mathbf{C}_{A,ij}$.

\section{Results and discussion}

We tested the preconditioned nonlinear conjugate gradient method for soft magnetic thin films (micromagnetic standard problem 1 \cite{mcmichael1997micromagnetic}), soft magnetic cubes (micromagnetic standard problem 3 \cite{rave1998magnetic,hertel2002finite}) and permanent magnet materials \cite{fischbacher2017nonlinear}. 

\subsection{Implementation details}
The demagnetizing field is calculated from a magnetic scalar potential. 
The linear system for the scalar potential is solved using the conjugate gradient method with an algebraic multigrid preconditioner \cite{demidov2012modification}. 
We enclose the magnet with an air box, in order to treat the magnetostatic boundary value problem \cite{chen1997review}. For computing sparse matrix multiplications, vector updates, and dot products, 
an OpenCL library  with C++ binding \cite{demidov2013programming} was used. 
The tests were run on NVIDIA K80 cards. We used two graphic processor 
units (GPUs) in parallel if the problem was too big to fit in the 
memory (12~GB) of one GPU. When we report computation times we do not 
include the time to assemble the system matrices in the precomputation.

To compute hysteresis we change the field in small steps. At each field a minimum of the energy was computed by Algorithm \ref{alg4}. 
The iterations were stopped when the change in energy, the change in the magnetization, and the norm of the gradient were sufficiently small according to \cite{gill1981practical} 
\begin{eqnarray}
\label{eq:t1}
F(\mathbf{m}_{j}) - F(\mathbf{m}_{j+1}) & < & \tau_{F} ( 1+|F(\mathbf{m}_{j+1})| ), \\
\label{eq:t2}
|\mathbf{m}_{j} - \mathbf{m}_{j+1}|_{\infty} & < & \sqrt{\tau_{F}} ( 1+|\mathbf{m}_{j+1}|_{\infty} ), \\
\label{eq:t3}
|\mathbf{g}_{j+1}|_{\infty} & < & \sqrt[3]{\tau_{F}} ( 1+|F(\mathbf{m}_{j+1})| ).
\end{eqnarray}
Unless otherwise stated, we set the function tolerance in the stopping conditions (\ref{eq:t1}) to (\ref{eq:t3})  to $\tau_{F} = 10^{-10}$. \cite{koehler1992finite, fischbacher2017nonlinear}
Different stopping criteria were used for the linear conjugate gradient method when solving the preconditioning system. The number of maximum iterations is set to a small number. 
The iterations of Algorithm \ref{alg1} were stopped when one of the following criteria was fulfilled: 
\begin{eqnarray}
\label{eq:li1}
 j & > & j_\mathrm{max}, \\
\label{eq:li2}
 \mathbf{d}_{j}^{\mathrm T} \mathbf{C}\mathbf{d}_{j} & < &0, \\
\label{eq:li3}
|\mathbf{g}_{j+1}| &  < & \min(0.5,\sqrt{|\mathbf{b}|})|\mathbf{g}_{0}|.
\end{eqnarray}

Typically criterion (\ref{eq:li1}) causes an exit from the loop if the maximum number of iterations, $j_\mathrm{max}$, is smaller than $15$.

\begin{figure}
\centering
\includegraphics{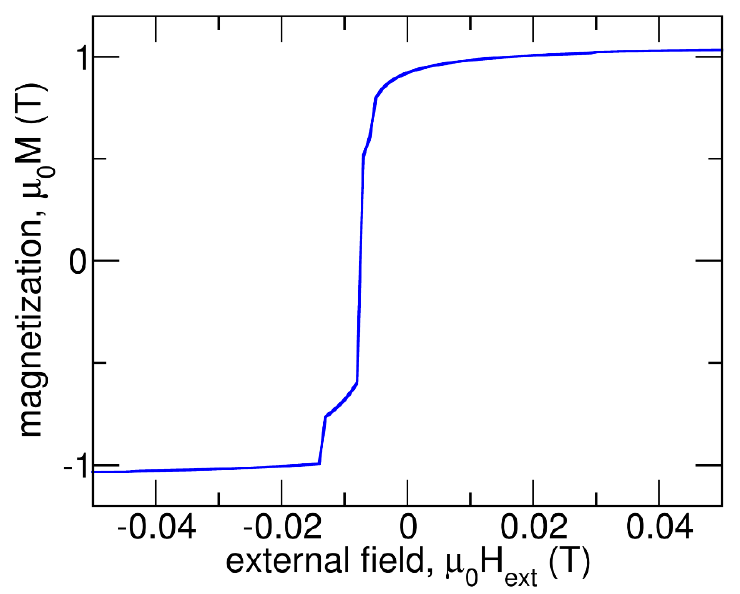}
\caption{\label{fig:std1} Demagnetization curve computed for the thin film element of the micromagnetic standard problem 1.}
\end{figure}

\setlength{\tabcolsep}{12pt}

\begin{table}
\caption{\label{tab:soft} Number of iterations and compute time for the 
micromagnetic standard problem number~1 with a mesh size of 5~nm. The 
first column gives the maximum number of iterations for solving the 
preconditioner system ($j_\mathrm{max}$) which is a parameter of the Algorithm. The other columns give the number of function evaluations ($evals$), the  number of nonlinear iterations ($nli$), the number of linear iterations ($li$), linear iterations for magnetostatics ($mi$), the total time for minimization ($nlt$), the time for the preconditioner ($lt$), and the time for magnetostatics ($mt$). The test runs were done on one GPU.}
\begin{center}
\begin{tabular}{r r r r r r r r}
\hline \hline
$j_\mathrm{max}$	&          $evals$& $nli$  & $li$ &       $mi$ &      $nlt$~(min) & $lt$~(min) & $mt$~(min) \\ \hline
          0       &            59015 & 29457  &      0 & 1739567 &          853 &        0 & 795 \\
          2       &            23173 & 11536  &  23274 &  698978 &          355 &       22 & 319 \\ 
  \textbf{4}      &   \textbf{19595} &   9747 &  39372 &  591830 & \textbf{312} &       24 & 270 \\ 
          6       &            22061 & 10980  &  66412 &  667845 &          363 &       20 & 304 \\
          8       &            23757 & 11828  &  94676 &  715704 &          401 &       22 & 326 \\
       1000       &            57445 & 28672  & 464094 & 1729425 &         1102 &      263 & 786 \\ \hline  \hline
\end{tabular}
\end{center}
\end{table}

\begin{table}
\caption{\label{tab:soft2} Number of iterations and compute time for 
the micromagnetic standard problem number~1 with a mesh size of 2.5~nm. 
See Table~\ref{tab:soft} for a description of the column headings. The 
test runs were done on two GPUs. }
\begin{center}
\begin{tabular}{r r r r r r r r}
\hline \hline
$j_\mathrm{max}$	&          $evals$& $nli$  & $li$ &       $mi$ &         $nlt$~(min) & $lt$~(min) & $mt$~(min) \\ \hline
          4       &            33635   &   16767 &   67472  &   1523277 &   2503 &        188 & 2180 \\ 
 \textbf{10}       &   \textbf{18107}  &    9003 &   90774  &  835459  &  \textbf{1579} &       263 & 1239 \\ \hline  \hline
\end{tabular}
\end{center}
\end{table}

\subsection{Micromagnetic standard problem 1}

The test case for soft magnetic materials was the micromagnetic standard problem number 1 \cite{mcmichael1997micromagnetic}. The sample is a permalloy film with dimensions
$1000 \times 2000 \times 20$~nm$^{3}$. The material parameters are a magnetization of $\mu_{0}M_{\mathrm s} = 1.05$~T and an exchange
constant of $A = 13$~pJ/m. There is uniaxial magneto-crystalline anisotropy along the long axis of the
sample with an anisotropy constant of $K = 500$~J/m$^{3}$. We compute the demagnetization curve with
the external field applied at one degree with respect to the long axis of the sample. For the soft magnetic test the external field was changed from $\mu_{0} H_\mathrm{ext,start} = 0.05$~T to $\mu_{0} H_\mathrm{ext,stop} = -0.05$~T in steps of $\mu_{0} \Delta H_\mathrm{ext} = 0.001$~T. The mesh size was 5~nm, which corresponds to $0.92 l_\mathrm{ex}$.  The exchange length, $l_\mathrm{ex}$, is defined as $l_\mathrm{ex} = \sqrt{2A/(\mu_{0}M_{\mathrm s}^{2})}$.  The total number of active degrees of freedom is $2.3 \times 10^{6}$. This is the sum of the magnetization components at the nodes of the magnetic thin film and the magnetic scalar potential at the nodes of the magnet and in the air box that encloses the magnetic sample. 

Fig. \ref{fig:std1} shows the computed demagnetization curve. Table \ref{tab:soft} gives the number of iterations and the compute time for the different tasks. The  total number of function evaluations is $evals$, the total number of nonlinear conjugate gradient iterations is $nli$, the number of iterations for solving the preconditioning system is $li$, and the number of iterations for solving the linear system for the scalar potential is $mi$. Similarly, the time for the nonlinear conjugate gradient method $nlt$, the time for solving the preconditioning system is $lt$, and the time for solving the linear system for the scalar potential is $mt$. The results show that only a few iterations of Algorithm \ref{alg1} are required in order to achieve a speed-up owing
to preconditioning the nonlinear conjugate gradient method. Note that $j_\mathrm{max} = 0$ is the nonlinear conjugate gradient method without preconditioning.  For $j_\mathrm{max} = 4$ the total number of function evaluations and the time to solution decrease by a factor smaller than 1/2.

In order to check the convergence with respect to the size of the finite element mesh we repeated the calculations with a mesh size of 2.5~nm ($0.46 l_\mathrm{ex}$). The total number of finite elements was $40\times10^{6}$ with $20\times10^{6}$ active degrees of freedom. The time for computing the demagnetization curve was 26.3 hours for $j_\mathrm{max}=10$. Table~\ref{tab:soft2} shows the number of iterations and the time to solution for computing the demagnetization curve. Although the number of degrees of freedom increased by a factor of 10 as compared to the simulation with a mesh size of 5 nm, the number of function evaluations is almost the same for the optimal choice of the preconditioner. The demagnetization curve is identical to that shown in Fig.~\ref{fig:std1}.

\begin{figure}
\centering
\includegraphics{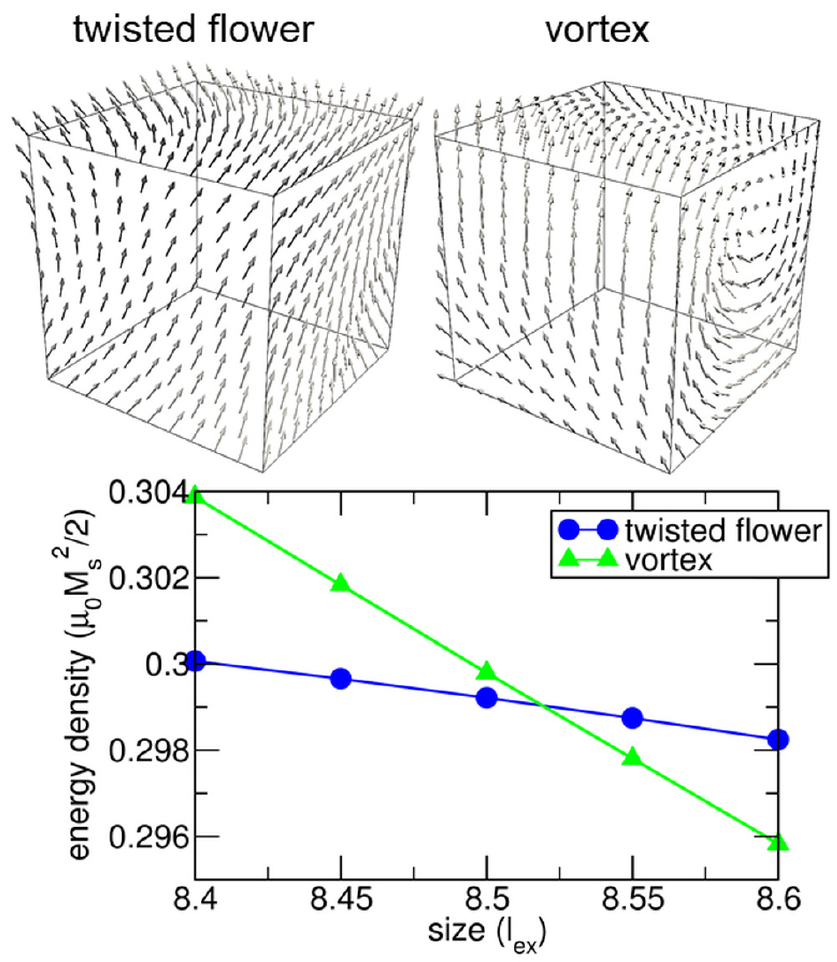}
\caption{\label{fig:std3} Energy densities of the the twisted flower state and the vortex state for the micromagnetic standard problem 3 as a function of the edge length of a cube. The edge length, $L$, is given in units of the exchange length, $l_\mathrm{ex}$. The top row shows the two different states for $L=8.5 l_\mathrm{ex}$}.
\end{figure}

\begin{table}
\caption{\label{tab:std3} Number of iterations and compute time for relaxing the vortex state for the micromagnetic standard problem number 3. See Table~\ref{tab:soft} for a description of the column headings. The test runs were done on one GPU.}
\begin{center}
\begin{tabular}{r r r r r r r r}
\hline \hline
$j_\mathrm{max}$	&          $evals$& $nli$  & $li$ &       $mi$ &      $nlt$~(s) & $lt$~(s) & $mt$~(s) \\ \hline
          0       &            603 &   301  &      0 &     17027 &      51.9 &          0 & 41.0 \\
          4       &            179 &    88  &    337 &      5174 &      19.4 &        4.0 & 12.5 \\ 
          8       &            127 &    61  &    420 &      3758 &      17.0 &        5.4 & 9.3 \\ 
 \textbf{12}       &            107 &    51  &    484 &      3177 &     \textbf{15.8} &        6.2 & 7.8 \\
         16       &             99 &    47  &    564 &      2972 &      16.3 &        7.2 & 7.3 \\ 
         20       &    \textbf{97} &    46  &    672 &      2916 &      17.3 &        8.5 & 7.1 \\
         24       &            107 &    51  &    906 &      3211 &      21.1 &        11.4 & 7.9 \\ 
      1000        &            303 &   149  &    4776 &     8911 &      85.7 &        58.9 & 21.6 \\ \hline \hline
\end{tabular}
\end{center}
\end{table}

\subsection{Micromagnetic standard problem 3}

The micromagnetic standard problem 3 is to calculate the single domain 
limit of a soft magnetic cube. For the simulation we used values for 
$M_ \mathrm{s}$ and $A$ similar to those of the micromagnetic standard problem 1. The anisotropy constant was $K = 0.1 \times \mu_{0} M_\mathrm{s}^{2}/2.$ The mesh size was $0.35 l_\mathrm{ex}$. In order to compute the energy of the single domain state and the multidomain state we relaxed the system either starting from the saturated state or a two-domain state with a perfectly sharp transition. We stopped the simulation when the criteria (\ref{eq:t1}) to (\ref{eq:t3}) were fulfilled with $\tau_{F} = 10^{-12}$. For cubes in the size range of $8.4 l_\mathrm{ex}$ to $8.6 l_\mathrm{ex}$ the two final states found were the twisted flower state and the vortex state. The twisted flower state and the vortex state (see Fig.~\ref{fig:std3}) were obtained when starting from the saturated state and the two-domain state, respectively. We computed the energy of the two states as function of the size of the cube. The critical size at which the twisted flower state and the vortex state have the same energy is $8.52 l_\mathrm{ex}$ which is close the the value reported by Hertel and Kronm\"uller \cite{hertel2002finite} ($8.56 l_\mathrm{ex}$). Though Hertel and Kronm\"uller applied a conjugate gradient method to minimize the micromagnetic energy, their finite element micromagnetic solver is different from our implementation. They used polar coordinates to satisfy the unit norm constraint and computed the magnetic vector potential to evaluate the magnetostatic energy. 

Again preconditioning sped up the simulations. Table~\ref{tab:std3} shows the iterations and the compute time for relaxing the vortex state for a size of 8.5 $l_\mathrm{ex}$. Preconditioning gives a speed up by a factor of 3.

\begin{table}
\caption{\label{tab:hard} Number of iterations and compute time for permanent magnet materials. See Table~\ref{tab:soft} for a description of the column headings. The test runs were done on one GPU.}
\begin{center}
\begin{tabular}{r r r r r r r r}
\hline \hline
$j_\mathrm{max}$	&          $evals$ &  $nli$  &   $li$ &       $mi$ &      $nlt$~(min) & $lt$~(min) & $mt$~(min) \\ \hline
          0       &            34673 &   16966 &      0 &     942511 &          709 &      0 &  661 \\
          2       &            12165 &    5712 &  12948 &     371348 &          287 &     13 &  258 \\ 
          4       &             8950 &    4104 &  18852 &     272754 &          222 &     20 &  189 \\ 
          6       &             8064 &    3661 &  24827 &     250678 &          212 &     25 &  176 \\
          8       &             6974 &    3116 &  28122 &     221311 &          191 &     27 &  155 \\
 \textbf{10}      &    \textbf{6585} &    2922 &  32813 &     210275 & \textbf{186} &     31 &  146 \\
         12       &             6583 &    2921 &  38978 &     209981 &          191 &     37 &  146 \\
         14       &             6641 &    2949 &  45463 &     211249 &          199 &     43 &  147 \\
         16       &             6834 &    3045 &  51804 &     216784 &          209 &     49 &  151 \\
       1000       &             7486 &    3357 &  94846 &     235638 &          263 &     90 &  164 \\ \hline  \hline
\end{tabular}
\end{center}
\end{table}

\subsection{Permanent magnets}

We used two different magnetic structures for testing the algorithm for 
the micromagnetic simulation of permanent magnets. The first test case 
for permanent magnets was similar to that introduced by Fischbacher and co-workers \cite{fischbacher2017nonlinear}. The sample is made of one layer of 9 grains. Each grain has an aspect ratio of approximately 2:2:1.  The extension of the magnet is about $460 \times 460 \times 100$~nm$^{3}$. The grains are separated by a non-magnetic grain boundary phase with a thickness of 5 nm.  The intrinsic magnetic properties of the hard magnetic phase were $K = 2.37$~MJ/m$^{3}$, saturation
magnetization $\mu_{0}M_{\mathrm s} = 1.61$~T, and exchange constant $A = 9.2$~pJ/m. The total number of degrees of freedom for this test was $6.9 \times 10^{6}$. 
For computing the demagnetization curve we started with an external field $\mu_{0}H_\mathrm{ext,start} = 0$. 
Then we decreased the external field in steps of  $\mu_{0} \Delta H_\mathrm{ext}= 0.005$~T. The final field value was $\mu_{0}H_\mathrm{ext,stop} = -3.7$~T. 
For this sample the proposed preconditioner for the nonlinear conjugate 
gradient method reduces the time to solution to about 1/4 of the time required to compute the demagnetization curve without preconditioning (see Table \ref{tab:hard}). The minimum number of function evaluations occurs when the linear system for preconditioning is solved with 10 iterations of the diagonally scaled conjugate gradient method (Algorithm \ref{alg1}). For every outer iteration of the nonlinear conjugate gradient method about 10 inner iterations for the linear system solver are performed. Since we drop the magnetostatic interaction for the inner iterations the preconditioner accounts for 16 percent of the total compute time. 

\begin{figure}
\centering
\includegraphics{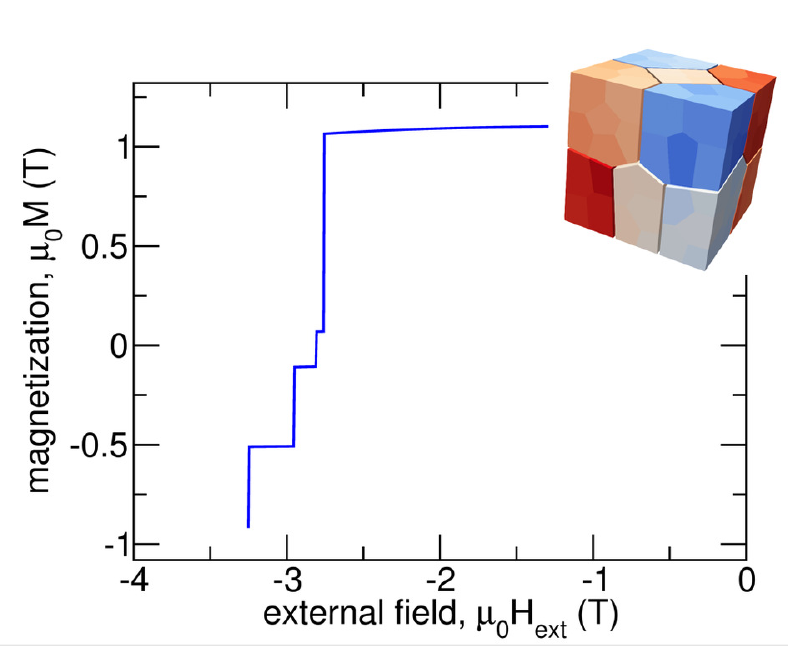}
\caption{\label{fig:grains} Computed demagnetization curve of a nanocrystalline permanent magnet.}
\end{figure}

\begin{table}
\caption{\label{tab:hard2} Number of iterations and compute time  for computing the demagnetization  curve of a  nanocrystalline permanent magnet (Fig.~\ref{fig:grains}). See Table~\ref{tab:soft} for a description of the column headings. The test runs were done on two GPUs.}
\begin{center}
\begin{tabular}{r r r r r r r r}
\hline \hline
$j_\mathrm{max}$	&          $evals$ &  $nli$  &   $li$ &       $mi$ &      $nlt$~(min) & $lt$~(min) & $mt$~(min) \\ \hline
          0       &           209595 &  104471 &      0 &     4490779 &          4359 &         0 &  3860 \\
         10       &             23169 &    7062 &  73323 &     455486 &          562 &       118 &   397 \\ \hline  \hline
\end{tabular}
\end{center}
\end{table}

The second test case for permanent magnets was a nanocrystalline 
magnet. The model was made of 12 grains with an average grain size of 
110~nm. The grains are separated by a 4~nm thick grain boundary phase. 
The magnetization of the grain boundary phase was $\mu_{0}M_\mathrm{s} = 0.54$~T. The average misorientation angle of the grains was 8 degrees. Each grain is divided into 10 patches. From patch to patch the anisotropy constant varies which mimics local changes in the chemical composition of the magnet. The magnetocrystalline anisotropy constants were Gaussian distributed with a mean of $K = 4.15$~MJ/m$^{3}$ and a standard deviation of 5 percent. The saturation magnetization $\mu_{0}M_\mathrm{s} = 1.19$~T and exchange constant $A = 10$~pJ/m. Fig.~\ref{fig:grains} shows the grain structure and the computed demagnetization curve. The number of active degrees of freedom was $29.2 \times 10^{6}$. The number of tetrahedrons was $16.8 \times 10^{6}$. The time to compute the curve shown in Fig.~\ref{fig:grains} was 72.6 hours without preconditioning. This was reduced to 9.4 hours by applying the preconditioner with the parameter $j_\mathrm{max} =10$.  The iteration counts of Table~\ref{tab:hard2} show that preconditioning reduces the number of nonlinear conjugate gradient iterations by a factor of 1/14. 

\subsection{Discussion}
For both the soft magnetic thin film and the hard magnetic granular structure only a few conjugate gradient iterations are required in order to improve the convergence of the nonlinear conjugate gradient method. 
Interestingly, preconditioning becomes worse when the preconditioning system is solved too accurately. This indicates that the local quadratic model of the preconditioner, $(1/2)\mathbf{y}^{\mathrm T}\mathbf{P}\mathbf{y}$,  is valid only in the close vicinity of the current approximation $\mathbf{x}_{j}$. We start Algorithm \ref{alg1} for solving $\mathbf{P} \mathbf{y} = \mathbf{g}$ with the initial guess $\mathbf{y}=0$. With each iteration of the linear conjugate gradient method  $|\mathbf{y}|$ growths \cite{steihaug1983conjugate}. We might reach a point which is outside the region where the quadratic model is a good approximation of the micromagnetic energy. Therefore,
we need to exit the linear conjugate gradient iterations after a few iterations. Nevertheless, the numerical tests showed significant reduction of the time to solution.

Moreover, we examined the influence of increasing system size to the necessary amount of iterations for the preconditioner system and overall function evaluations. 
Numerical tests indicate that the optimal number of iterations for preconditioning increases slightly if the discretization is refined. On the other hand, 
 the overall number of function evaluations stay approximately the same,  which is in agreement with the usually expected behavior for successfully preconditioned methods like linear conjugate gradients. 
 That is, the number of iterations of our proposed method does not increase with the system size. As a consequence the algorithm is nearly optimal. The time to solution scales almost linearly with  problem size. Let us have a look at the computing times for the micromagnetic standard problem~1. Please note that the number of unknowns are the 
 Cartesian components of the unit magnetization vector in the magnet and the magnetic scalar potential in the magnet and the air box. For $2.3 \times 10^{6}$ degrees of freedom computing the demagnetization curve took 5.2~hours on one GPU. For $20 \times 10^{6}$ degrees of freedom the time to solution was 26.3~hours on two GPUs. 
 Though a detailed study of the complexity of the algorithm would require more test runs we see the following. Ideally, increasing the system size by factor of 9 (from $2.3 \times 10^{6}$ to $20 \times 10^{6}$ degrees of freedom) and doubling the compute power should increase the time to solution by a factor of 4.5. The actual ratio between the compute times for the two different problem sizes is 5.06. 
 The complexity of the function evaluation is mostly due to the solution of the magnetostatic subproblem. The combination of a linearly scaling method for the magnetostatic problem for which we use an algebraic multigrid 
 method together with successful preconditioning led to nearly optimal scaling.

\section{Conclusion}
A nonlinear conjugate gradient method has been developed which successfully tackles the computationally demanding nonlinear and non-convex constrained optimization problem in micromagnetics. 
 We presented an efficient preconditioning strategy for the nonlinear conjugate gradient method in micromagnetics 
with significant speed up as demonstrated in several test cases.  The preconditioning approach only requires a few iterations (up to 10) of a linear conjugate gradient solver involving only sparse matrix vector 
products. However, as for the well-known linear conjugate gradient method for unconstrained minimization of quadratic functionals, preconditioning the nonlinear conjugate gradient method 
aims at maintaining the (almost) linear scaling independent of increasing problem size. Our tests on several (nontrivial) problems computed on GPUs indicate exactly this behavior.

\section*{Acknowledgments}
Financial support by the Austrian Science Fund (FWF) via the SFB "ViCoM" under grant No. F41 and SFB "Complexity in PDEs" under grant No. F65 is acknowledged. 

\bibliographystyle{abbrv} 
\bibliography{prewithtitle}

\providecommand{\noopsort}[1]{}\providecommand{\singleletter}[1]{#1}%
\begin{thebibliography}{10}

\bibitem{abert2014efficient}
C.~Abert, G.~Wautischer, F.~Bruckner, A.~Satz, and D.~Suess.
\newblock Efficient energy minimization in finite-difference micromagnetics:
  {S}peeding up hysteresis computations.
\newblock {\em Journal of Applied Physics}, 116(12):123908, 2014.

\bibitem{acevedo2016efficient}
M.~A.~E. Acevedo.
\newblock Efficient micromagnetics for magnetic storage devices.
\newblock University of California, San Diego, 2016.

\bibitem{andrea2017novel}
C.~Andrea, F.~Giovanni, and R.~Massimo.
\newblock Novel preconditioners based on quasi--{N}ewton updates for nonlinear
  conjugate gradient methods.
\newblock {\em Optimization Letters}, 11(4):835--853, 2017.

\bibitem{apalkov2003fast}
D.~Apalkov and P.~Visscher.
\newblock Fast multipole method for micromagnetic simulation of periodic
  systems.
\newblock {\em IEEE Trans. Magn.}, 39(6):3478--3480, 2003.

\bibitem{bance2014high}
S.~Bance, H.~Oezelt, T.~Schrefl, M.~Winklhofer, G.~Hrkac, G.~Zimanyi,
  O.~Gutfleisch, R.~Evans, R.~Chantrell, T.~Shoji, et~al.
\newblock High energy product in {B}attenberg structured magnets.
\newblock {\em Applied Physics Letters}, 105(19):192401, 2014.

\bibitem{brueckl2017vortex}
H.~Brueckl, A.~Satz, K.~Pruegl, T.~Wurft, S.~Luber, W.~Raberg, J.~Zimmer, and
  D.~Suess.
\newblock Vortex magnetization state in a {GMR} spin-valve type field sensor.
\newblock In {\em Magnetics Conference (INTERMAG), 2017 IEEE International},
  pages 1--1. IEEE, 2017.

\bibitem{chen1997review}
Q.~Chen and A.~Konrad.
\newblock A review of finite element open boundary techniques for static and
  quasi-static electromagnetic field problems.
\newblock {\em IEEE Transactions on Magnetics}, 33(1):663--676, 1997.

\bibitem{cohen1989relaxation}
R.~Cohen, S.-Y. Lin, and M.~Luskin.
\newblock Relaxation and gradient methods for molecular orientation in liquid
  crystals.
\newblock {\em Computer Physics Communications}, 53(1-3):455--465, 1989.

\bibitem{della1993stiff}
E.~Della~Torre.
\newblock Stiff nodes in numerical micromagnetic problems.
\newblock {\em IEEE transactions on magnetics}, 29(6):2371--2373, 1993.

\bibitem{demidov2013programming}
D.~Demidov, K.~Ahnert, K.~Rupp, and P.~Gottschling.
\newblock Programming {CUDA} and {OpenCL}: {A} case study using modern {C}++
  libraries.
\newblock {\em SIAM Journal on Scientific Computing}, 35(5):C453--C472, 2013.

\bibitem{demidov2012modification}
D.~Demidov and D.~Shevchenko.
\newblock Modification of algebraic multigrid for effective {GPU}-based
  solution of nonstationary hydrodynamics problems.
\newblock {\em Journal of Computational Science}, 3(6):460--462, 2012.

\bibitem{exl2014thesis}
L.~Exl.
\newblock Tensor grid methods for micromagnetic simulations.
\newblock Vienna UT (thesis), 2014.

\bibitem{exl2014labonte}
L.~Exl, S.~Bance, F.~Reichel, T.~Schrefl, H.~Peter~Stimming, and N.~J. Mauser.
\newblock La{B}onte's method revisited: {A}n effective steepest descent method
  for micromagnetic energy minimization.
\newblock {\em Journal of Applied Physics}, 115(17):17D118, 2014.

\bibitem{exl2017extrapolated}
L.~Exl, N.~J. Mauser, T.~Schrefl, and D.~Suess.
\newblock The extrapolated explicit midpoint scheme for variable order and step
  size controlled integration of the {L}andau-{L}ifschitz-{G}ilbert equation.
\newblock {\em Journal of Computational Physics}, 346:14–24, 2017.

\bibitem{exl2014non}
L.~Exl and T.~Schrefl.
\newblock Non-uniform {FFT} for the finite element computation of the
  micromagnetic scalar potential.
\newblock {\em J. Comput. Phys.}, 270:490--505, 2014.

\bibitem{fabian1996three}
K.~Fabian, A.~Kirchner, W.~Williams, F.~Heider, T.~Leibl, and A.~Huber.
\newblock Three-dimensional micromagnetic calculations for magnetite using
  {FFT}.
\newblock {\em ‎Geophys. J. Int.}, 124(1):89--104, 1996.

\bibitem{fischbacher2017limits}
J.~Fischbacher, A.~Kovacs, H.~Oezelt, M.~Gusenbauer, T.~Schrefl, L.~Exl,
  D.~Givord, N.~Dempsey, G.~Zimanyi, M.~Winklhofer, et~al.
\newblock On the limits of coercivity in permanent magnets.
\newblock {\em Applied Physics Letters}, 111(7):072404, 2017.

\bibitem{fischbacher2017nonlinear}
J.~Fischbacher, A.~Kovacs, H.~Oezelt, T.~Schrefl, L.~Exl, J.~Fidler, D.~Suess,
  N.~Sakuma, M.~Yano, A.~Kato, et~al.
\newblock Nonlinear conjugate gradient methods in micromagnetics.
\newblock {\em AIP Advances}, 7(4):045310, 2017.

\bibitem{fletcher1964function}
R.~Fletcher and C.~M. Reeves.
\newblock Function minimization by conjugate gradients.
\newblock {\em The computer journal}, 7(2):149--154, 1964.

\bibitem{fu2015micromagnetics}
S.~Fu, R.~Chang, S.~Couture, M.~Menarini, M.~Escobar, M.~Kuteifan, M.~Lubarda,
  D.~Gabay, and V.~Lomakin.
\newblock Micromagnetics on high-performance workstation and mobile
  computational platforms.
\newblock {\em J. Appl. Phys.}, 117(17):17E517, 2015.

\bibitem{garcia2008numerical}
C.~Garc{\'\i}a-Cervera.
\newblock Numerical micromagnetics: A review.
\newblock {\em Bolet{\'\i}n SEMA}, (39):103--135, 2008.

\bibitem{gill1981practical}
P.~E. Gill, W.~Murray, and M.~H. Wright.
\newblock {\em Practical optimization}.
\newblock Academic press, 1981.

\bibitem{goldfarb2009curvilinear}
D.~Goldfarb, Z.~Wen, and W.~Yin.
\newblock A curvilinear search method for p-harmonic flows on spheres.
\newblock {\em SIAM Journal on Imaging Sciences}, 2(1):84--109, 2009.

\bibitem{hertel2002finite}
R.~Hertel and H.~Kronm{\"u}ller.
\newblock Finite element calculations on the single-domain limit of a
  ferromagnetic cube—a solution to $\mu$mag {S}tandard {P}roblem {N}o. 3.
\newblock {\em Journal of Magnetism and Magnetic Materials}, 238(2):185--199,
  2002.

\bibitem{hestenes1952methods}
M.~Hestenes.
\newblock Methods of conjugate gradients for solving linear systems.
\newblock {\em J. Res. Nat. Bur. Standards}, 49:409--435, 1952.

\bibitem{kinderlehrer1994simulation}
D.~S. Kinderlehrer and L.~Ma.
\newblock Simulation of hysteresis in nonlinear systems.
\newblock In {\em 1994 North American Conference on Smart Structures and
  Materials}, pages 78--87. International Society for Optics and Photonics,
  1994.

\bibitem{koehler1992finite}
T.~Koehler and D.~Fredkin.
\newblock Finite element methods for micromagnetics.
\newblock {\em IEEE transactions on magnetics}, 28(2):1239--1244, 1992.

\bibitem{kovacs2017micromagnetic}
A.~Kovacs, H.~Ozelt, J.~Fischbacher, T.~Schrefl, A.~Kaidatzis, R.~Salikhof,
  M.~Farle, G.~Giannopoulos, and D.~Niarchos.
\newblock Micromagnetic {S}imulations for {C}oercivity {I}mprovement through
  {N}ano-{S}tructuring of {R}are-{E}arth {F}ree {L1 0-FeNi} {M}agnets.
\newblock {\em IEEE Transactions on Magnetics}, 2017.

\bibitem{kritsikis2014beyond}
E.~Kritsikis, A.~Vaysset, L.~Buda-Prejbeanu, F.~Alouges, and J.-C. Toussaint.
\newblock Beyond first-order finite element schemes in micromagnetics.
\newblock {\em J. Comput. Phys.}, 256:357--366, 2014.

\bibitem{mcmichael1997micromagnetic}
R.~D. McMichael and M.~J. Donahue.
\newblock Micromagnetic computational standard problem.
\newblock {\em Journal of Applied Physics}, 81(8):5242--5242, 1997.

\bibitem{nocedal2006numerical}
J.~Nocedal and S.~Wright.
\newblock Numerical optimization.
\newblock {\em New York}, 2006.

\bibitem{oikawa2016large}
T.~Oikawa, H.~Yokota, T.~Ohkubo, and K.~Hono.
\newblock Large-scale micromagnetic simulation of {Nd-Fe-B} sintered magnets
  with {Dy}-rich shell structures.
\newblock {\em AIP Advances}, 6(5):056006, 2016.

\bibitem{palmesi2017high}
P.~Palmesi, L.~Exl, F.~Bruckner, C.~Abert, and D.~Suess.
\newblock Highly parallel demagnetization field calculation using the fast
  multipole method on tetrahedral meshes with continuous sources.
\newblock {\em Journal of Magnetism and Magnetic Materials}, 442:409–416,
  2017.

\bibitem{polak1969note}
E.~Polak and G.~Ribiere.
\newblock Note sur la convergence de m{\'e}thodes de directions conjugu{\'e}es.
\newblock {\em Revue fran{\c{c}}aise d'informatique et de recherche
  op{\'e}rationnelle. S{\'e}rie rouge}, 3(16):35--43, 1969.

\bibitem{polyak1969conjugate}
B.~T. Polyak.
\newblock The conjugate gradient method in extremal problems.
\newblock {\em USSR Computational Mathematics and Mathematical Physics},
  9(4):94--112, 1969.

\bibitem{popovic2005applications}
N.~Popovi{\'c} and D.~Praetorius.
\newblock Applications of {H}-{M}atrix techniques in micromagnetics.
\newblock {\em Computing}, 74(3):177--204, 2005.

\bibitem{pytlak2008conjugate}
R.~Pytlak.
\newblock {\em Conjugate gradient algorithms in nonconvex optimization},
  volume~89.
\newblock Springer Science \& Business Media, 2008.

\bibitem{rave1998magnetic}
W.~Rave, K.~Fabian, and A.~Hubert.
\newblock Magnetic states of small cubic particles with uniaxial anisotropy.
\newblock {\em Journal of Magnetism and Magnetic Materials}, 190(3):332--348,
  1998.

\bibitem{salleh2016efficient}
Z.~Salleh and A.~Alhawarat.
\newblock An efficient modification of the {H}estenes-{S}tiefel nonlinear
  conjugate gradient method with restart property.
\newblock {\em Journal of Inequalities and Applications}, 2016(1):110, 2016.

\bibitem{schrefl_numerical_nodate}
T.~Schrefl, G.~Hrkac, S.~Bance, D.~Suess, O.~Ertl, and J.~Fidler.
\newblock Numerical {M}ethods in {M}icromagnetics ({FEMS}).
\newblock In H.~Kron\"uller and S.~Parkin, editors, {\em Handbook of Magnetism
  and Advanced Magnetic Materials}, volume~2. Wiley, 2007.

\bibitem{sepehri2016micromagnetic}
H.~Sepehri-Amin, T.~Ohkubo, and K.~Hono.
\newblock Micromagnetic simulations of magnetization reversals in {Nd-Fe-B}
  based permanent magnets.
\newblock {\em MATERIALS TRANSACTIONS}, 57(8):1221--1229, 2016.

\bibitem{shepherd2014discretization}
D.~Shepherd, J.~Miles, M.~Heil, and M.~Mihajlovi{\'c}.
\newblock Discretization-induced stiffness in micromagnetic simulations.
\newblock {\em IEEE Transactions on Magnetics}, 50(11):1--4, 2014.

\bibitem{sodervznik2017magnetization}
M.~Soder{\v{z}}nik, H.~Sepehri-Amin, T.~Sasaki, T.~Ohkubo, Y.~Takada, T.~Sato,
  Y.~Kaneko, A.~Kato, T.~Schrefl, and K.~Hono.
\newblock Magnetization reversal of exchange-coupled and exchange-decoupled
  {Nd-Fe-B} magnets observed by magneto-optical kerr effect microscopy.
\newblock {\em Acta Materialia}, 2017.

\bibitem{steihaug1983conjugate}
T.~Steihaug.
\newblock The conjugate gradient method and trust regions in large scale
  optimization.
\newblock {\em SIAM Journal on Numerical Analysis}, 20(3):626--637, 1983.

\bibitem{suess2002time}
D.~Suess, V.~Tsiantos, T.~Schrefl, J.~Fidler, W.~Scholz, H.~Forster,
  R.~Dittrich, and J.~Miles.
\newblock Time resolved micromagnetics using a preconditioned time integration
  method.
\newblock {\em Journal of Magnetism and Magnetic Materials}, 248(2):298--311,
  2002.

\bibitem{sun2006adaptive}
J.~Sun and P.~Monk.
\newblock An adaptive algebraic multigrid algorithm for micromagnetism.
\newblock {\em IEEE Trans. Magn.}, 42(6):1643--1647, 2006.

\bibitem{tanaka2017speeding}
T.~Tanaka, A.~Furuya, Y.~Uehara, K.~Shimizu, J.~Fujisaki, T.~Ataka, and
  H.~Oshima.
\newblock Speeding up micromagnetic simulation by energy minimization with
  interpolation of magnetostatic field.
\newblock {\em IEEE Transactions on Magnetics}, 2017.

\bibitem{tsiantos2006fast}
V.~Tsiantos and J.~Miles.
\newblock Fast micromagnetic simulations using an analytic mathematical model.
\newblock {\em Physica B: Condensed Matter}, 372(1):303--307, 2006.

\bibitem{tsukahara2016implementation}
H.~Tsukahara, K.~Iwano, C.~Mitsumata, T.~Ishikawa, and K.~Ono.
\newblock Implementation of low communication frequency {3D FFT} algorithm for
  ultra-large-scale micromagnetics simulation.
\newblock {\em Computer Physics Communications}, 207:217--220, 2016.

\bibitem{yuan1992fast}
S.~W. Yuan and H.~N. Bertram.
\newblock Fast adaptive algorithms for micromagnetics.
\newblock {\em IEEE Trans. Magn.}, 28(5):2031--2036, 1992.

\end{thebibliography}

\end{document}